\documentclass[10pt,twoside]{article}
\usepackage{amsmath,amssymb}
\usepackage{Latex-document}

\DeclareMathOperator{\dd}{d}

\newcommand{\MM}{M}

\newcommand{\RR}{\mathbb{R}}
\newcommand{\SSS}{\mathbb{S}}

\newcommand{\pr}{\partial}

\newcommand{\nn}{\vec{n}}

\newcommand{\eps}{\varepsilon}
\newcommand{\ee}{e}

\newcommand{\AAA}{\mathcal{A}}

\def\volumeno{I}

\title{\bf Knotted Solitons\vskip 6mm}

\author{L. D. Faddeev\thanks{St.~Petersburg Department
of Steklov Mathematical Institute, Russian Academy of Sciences,
Russia. E-mail: faddeev@pdmi.ras.ru}\vspace*{-0.5cm}}

\date{\vspace{-8mm}}

\begin{document}

\maketitle

\thispagestyle{first} \setcounter{page}{235}

\begin{abstract}

\vskip 3mm

The dynamical model on 3+1 dimensional space-time admitting
soliton solutions is discussed. The proposal soliton is localized
in the vicinity of a closed contour, which could be linked and/or
knotted. The topological charge is Hopf invariant. Some
applications in realistic physical systems are indicated.

\vskip 4.5mm


\end{abstract}

\vskip 12mm

\section{Introduction}

\vskip-5mm \hspace{5mm}

        The term ``soliton'' entered applied mathematics in 1965.
        It was coined by M.~Kruskal and N.~Zabusky for a special
        solution of nonlinear Korteweg-de Vries (KdV) equation,
        depicting solitary wave
\cite{KZ}.
        Use of convention of particle physics language shows that
    the author
        envisioned the particle-like interpretation for the object
        which they called soliton.

        The attention of mathematical physicists to solitons was
        attracted after the inverse scattering method was devised
        by G.~Gardner, J.~Green, M.~Kruskal and R.~Miura for solving the
        KdV equation
\cite{ggkm}
    and its extension to Nonlinear Schroedinger
        Equation was found by V.~Zakharov and A.~Shabat
\cite{ZSh}.
        In the 1970's this method and its generalizations got a
        lot of attention and involved quite a few active participants.
    Rather complete review can be found in
\cite{FT}.
        In the end of that decade the quantum variant of the method
        was constructed and particle-like interpretations of solitons
        got natural confirmation in terms of quantum field theory,
    see review in
\cite{LH}.
        The mathematical structure of the quantum method was deciphered
        in pure algebraic way leading in the 1980's to notion of
        quantum groups with new applications in pure mathematics and
        mathematical physics.

        The value of solitons for the particle physics consists in the
        possibility of going beyond the paradigm of the perturbation
        theory.
        Indeed, soliton solutions correspond to full nonlinear
        equations and disappear in their linearized form.
        Characteristic for solitons is that they interact strongly if the
        excitations of the linearized fields interact weakly.
        Another attractive feature is the  appearance of elementary topological
        characteristics for solitons topological charges.

        This was understood already in the middle of
        1970's by several groups as
        I underlined in my lectures, when I was touring USA in
        1975 (see e.~g.
\cite{princeton}).
        However, all these tantalizing features of solitons had one very
        important drawback: the developed methods applied only in
        1 + 1 dimensional space-time.

        Naturally the search for 3 + 1 dimensional generalizations became
        eminent.
        General considerations showed that many features of 1 + 1
        dimensional systems, such as complete integrability and
        existence of exact many-particle solutions could not be
        generalized to 3 + 1 dimensions.
        However, the mere existence of ``one-particle'' soliton solutions
        was not excluded.
        One particular example was introduced by Skyrme  in a
        pioneer paper
\cite{skyrme}
    long before the soliton rush.
        Another example was proposed by G.~'t~Hooft and A.~Polyakov
        in 1975
\cite{HP}.
        In the following years their solutions got real applications in
        nuclear and high energy physics.

        In both examples the solitons are ``point-like'', namely their
        deviation from the vacuum is concentrated around central point
        in space.
        Moreover they have spherical symmetry, allowing the separation
        of variables in the corresponding equations, reducing them
        to ODE, which one can treat on a usual PC.

        In my lectures
\cite{princeton},
    already mentioned, I proposed one more
        possibility for 3 + 1 dimensional system, allowing solitons.
        The model, which superficially looks as a slight modification
        of Skyrme model, has quite distinct features.
        The center of the would-be soliton is not a point, but a
        closed contour, possibly linked or knotted.
        However my proposal remained unnoticed.
        The reason was evident: the maximal symmetry for such a soliton
        is axial, reducing 3-dimensional nonlinear PDE to 2-dimensional
        one.
        Existing computers were not able to treat such a problem.
        Thus my proposal was in slumber for 20 years until my colleague
        Antti Niemi became interested and agreed to sacrifice a year to
        learn computing and devising the programm.
        The preliminary results published in
\cite{FNN}
    attracted the attention
        of professionals in computational physics and now we have an ample
        evidence, confirming my proposal
\cite{HS},
\cite{BS}.

        The development which followed showed unexpected universality of
        my model.
        The variables, used in it, were shown to enter the list of
        degrees of freedom for several systems, having realistic
        physical applications
\cite{FNP},
\cite{BFN}.

        In this talk I shall describe all these developments in detail.
        First I shall introduce the model, then briefly discuss its
        numerical treatment and finish with the description of the
        applications.

\section{The field configurations and Hopf invariant}

\vskip-5mm \hspace{5mm}

        The space time is 4-dimensional Minkowski space
    $ \MM $
        with linear coordinates
    $ x^{\mu} $,
    $ \mu=0,1,2,3 $,
    $ x^{0} $
        being time and
    $ x^{k} $,
    $ k=1,2,3 $
        space variables.
        The field
    $ \nn (x) $
        is defined on
    $ \MM $
        and has values on 2-dimensional sphere
    $ \SSS^{2} $:
\begin{equation*}
    \nn: \MM \to \SSS^{2} .
\end{equation*}
        The boundary condition on spatial infinity is introduced
\begin{equation}
\label{boundc}
    \nn |_{r=\infty} = \nn_{0} ,
\end{equation}
        where
    $ r = ((x^{1})^{2} + (x^{2})^{2} + (x^{3})^{2})^{1/2} $
        and
    $ \nn_{0} $
        is a fixed vector, e.g. corresponding to the north pole
\begin{equation*}
    \nn_{0} = (0,0,1) .
\end{equation*}
        We shall consider mostly the time independent configurations,
        corresponding to a soliton at rest.
        The boundary condition
(\ref{boundc})
    effectively compactifies the space
    $ \RR^{3} $,
        turning it into sphere
    $ \SSS^{3} $,
        thus the stationary configurations realize the map
\begin{equation}
\label{map23}
    \nn: \SSS^{3} \to \SSS^{2} ,
\end{equation}
        which are known to be classified by Hopf invariant, sort of
        topological charge.

        In general, the density of topological charge is the zero
        component
    $ J_{0} $
        of the current
    $ J_{\mu} $,
        which is conserved
\begin{equation*}
    \pr_{\mu} J_{\mu} = 0
\end{equation*}
        independently of the equations of motion.
        Mathematically it is more natural to use the 3-form
    $ J $
        dual to 1-form
    $ J^{*} = J_{\mu} \dd x^{\mu}$
        and define the topological charge as an integral of
    $ J $
        over space section
\begin{equation*}
    Q = \int_{\RR^{3}} J .
\end{equation*}

        In our case the 3-form
    $ J $
        is constructed as follows.
        The pull-back of the volume 2-form on
    $ \SSS^{2} $
        via map
(\ref{map23})
    defines the closed 2-form on the space time
\begin{equation*}
    H = H_{\mu\nu} \dd x^{\mu} \wedge \dd x^{\nu} ,
\end{equation*}
        where antisymmetric tensor
    $ H_{\mu\nu} $
        is expressed via field configuration
    $ \nn(x) $
        as follows
\begin{equation}
\label{HH}
    H_{\mu\nu} = (\pr_{\mu} \nn \times \pr_{\nu} \nn , \nn) .
\end{equation}
        Here I use usual notations of vector analysis in 3-space.
        In fact
    $ H $
        is exact
\begin{equation*}
    H = \dd C
\end{equation*}
        and current 3-form is given by
\begin{equation*}
    J = \frac{1}{4\pi} H \wedge C .
\end{equation*}
        In more detail, we have the relations
\begin{equation*}
    H_{ik} = \pr_{i} C_{k} - \pr_{k} C_{i}
\end{equation*}
        and
\begin{equation*}
    Q = \frac{1}{4\pi} \int \eps_{ikj} H_{ik} C_{j} d^{3}x .
\end{equation*}

        For regular configurations
    $ Q $
        gets integer values.
        This integer has a nice interpretation in the description of which
        I shall use the terminology of magnetostatic.

        Tensor
    $ H_{ik} $
        can be interpreted as a field strength of the stationary magnetic
        field in Maxwell theory.
        The corresponding lines of force are defined via equations
\begin{equation*}
    \frac{d}{ds} x_{i} = \frac{1}{2} \eps_{ikj} H_{kj} ,
\end{equation*}
        where
    $ s $
        is a local parameter along the line.
        It is easy to see that components of
    $ \nn(x) $
        along these lines are constant
\begin{equation*}
    \frac{d}{ds} \nn(x) = 0 ,
\end{equation*}
        giving two ``integrals of the motion''.
        In other words, the Maxwell lines of force are the preimages
        of points on
    $ \SSS^{2} $
        under the map
(\ref{map23}).
        Hopf invariant is the intersection number of any pair of such
        lines.

        All these facts are well known and can be found in textbooks
        (see e.g.
\cite{MM}).
        However I decided to include them into my text to make it
        more selfcontained.

\section{The dynamical model}

\vskip-5mm \hspace{5mm}

        I introduce the dynamical model by giving the relativistic
        action functional
\begin{equation*}
    \AAA = a \int (\pr_{\mu} \nn)^{2} d^{4}x +
        b \int (H_{\mu\nu})^{2} d^{4}x .
\end{equation*}
        In the usual convention of high-energy physics
    $ \AAA $
        is dimensionless, so the parameter
    $ a $
        has dimension
    $ [\text{length}]^{-2} $
        and parameter
    $ b $
        is dimensionless.
        Corresponding static energy
    $ E $
        has the same form as
    $ \AAA $
        with space-time coordinates substituted by space variables
        only
\begin{equation}
\label{energy}
    E = a \int (\pr_{k} \nn)^{2} d^{3}x +
        b \int (H_{ik})^{2} d^{3}x .
\end{equation}
        and has proper dimension
    $ [\text{length}]^{-1} $.
        The structure of
    $ E $
        is similar to that of Skyrme model, where the field variable
        having values in
    $ \SSS^{3} $
        is used and corresponding topological charge is just a degree
        of map.

        Usual check based on the scale transformation is favorable for
(\ref{energy})
        in the same way as in Skyrme model.
        Indeed
\begin{equation*}
    E = E_{2} + E_{4} ,
\end{equation*}
        where
    $  E_{2} $
        and
    $  E_{4} $
        are quadratic and quartic in derivatives of
    $ \nn $
        correspondingly.
        Thus under scaling
    $ x \to \lambda x $
        we have
\begin{equation*}
    E_{2} \to \lambda E_{2} \, , \quad
    E_{4} \to \frac{1}{\lambda} E_{4}
\end{equation*}
        and the virial theorem states that on the minimal
        configuration (if any)
\begin{equation*}
    E_{2} = E_{4}.
\end{equation*}

        In terms of quantum theory
    $  E_{2} $
        has a standard interpretation of the energy of nonlinear
        sigma-model whereas
    $  E_{4} $
        is rather exotic.
        On the contrary in the magnetic interpretation, mentioned above,
    $  E_{4} $
        is a natural term --- it is just the Maxwell magnetic energy,
        whereas the nature of
    $  E_{2} $
        is not that clear.
        However in what follows the presence of both
    $  E_{2} $
        and
    $  E_{4} $
        is crucial for the existence of solitons as the scaling argument
        already showed.

        This is confirmed also by a beautiful estimate, obtained in
\cite{VK}
\begin{equation*}
    E \geq c |Q|^{3/4} ,
\end{equation*}
        which shows that in the sectors with nonzero
    $ Q $
        the minimum of energy is strongly positive.
        Thus the soliton solutions should be obtained by the minimizing of
    $ E $
        with
    $ Q \neq 0 $
    fixed.

        Unfortunately until now there exists no proof of the compactness
        of the minimizing sequence in general case. For the case of axial
    symmetry uncouraging result are obtained in
\cite{RR}.
        So the main argument for the evidence of solitons in my model is
    based on the numerical work.

\section{Numerical work}

\vskip-5mm \hspace{5mm}

        To find the numerical evidence of the existence of localized
        solitons it is not necessary to solve the nonlinear elliptic
        equation, obtained by the variational principle
\begin{equation}
\label{ve}
    \frac{\delta E}{\delta \nn} = 0 .
\end{equation}
        Instead one can introduce an auxiliary time
    $ s $
        and consider the parabolic equation
\begin{equation}
\label{pe}
    \frac{d \nn}{ds} = \frac{\delta E}{\delta \nn}
\end{equation}
        with initial value
    $ \nn_{\text{init}} $
\begin{equation*}
    \nn |_{s=0} = \nn_{\text{init}}
\end{equation*}
        being a configuration with the prescribed Hopf invariant.
        Of course to simulate
(\ref{pe})
    on the computer one is to use some
        difference scheme.
        If for large
    $ s $
        solution of
(\ref{pe})
    stabilizes it gives the solution of
(\ref{ve}).
        In other words the soliton appears as an attractor for the
        evolution equation.

        There are of course many important practical details how to
        discretize equation,
    how to take into account the normalization condition
    $ \nn^{2} = 1$
        and how to choose the initial configuration
    $ \nn_{\text{init}} $.
        The main papers
\cite{HS}
        and
\cite{BS}
        use different prescription for all this, however quite
        satisfactorily the final results coincide.
        I refer to these papers for the details of calculations and
        proceed to describe the results.

        The iterative process was performed for the configuration with
    $ Q = 1,2,\ldots 7$.
        The results are as follows: for
    $ Q = 1 $
        and
    $ Q = 2 $
        the solutions are axial symmetric.
        The center line --- the preimage of the point
    $ n = (0,0,-1) $ ---
        is a circle.
        The surfaces
    $ n_{3} = \alpha $,
    $ -1 < \alpha < 1 $
        are toroidal and they are spanned by the lines of force
        wrapping the torus once for
    $ Q = 1 $
        and twice for
    $ Q = 2 $.
        In other words the soliton can be viewed as a filament of
        lines of force, closed and twisted once or twice.

        The solution for
    $ Q = 3 $
        is similar but not axial symmetric any more, the corresponding
        ``cable'' is warped.
        For
    $ Q = 4 $
        the soliton is a link of two twisted filaments.
        Especially beautiful case is
    $ Q = 7 $,
        the central line of the corresponding soliton is a trefoil knot.

        The file
\cite{HH}
        contains impressive moving pictures illustrating the convergence
        of the iterations.
        I plan to show these movies in my talk, but unfortunately can
        not do it in a written text.

        Thus the numerical work gives the compelling evidence of the
        existence of string-like solitons in my model.
        There remains an important mathematical challenge to provide
        the rigorous existence theorem.
        Another interesting direction is to find some realistic applications
        of the model.
        Some progress in this direction is already obtained and I
        proceed to the description of it.

\section{The applications}

\vskip-5mm \hspace{5mm}

        Nonlinear fields such as
    $ \nn(x) $
        rarely enter the dynamical models directly.
        However they can appear as a part of degrees of freedom in a
        suitable parameterization of the original fields.
        For example in condensed matter theory one uses the complex
        valued functions
    $ \psi_{\alpha}(x) $,
    $ \alpha = 1, \ldots , N $
        to describe the density amplitudes of Bose gas or the
        gap function of superconductor.
        The interaction supports the configurations, for which
\begin{equation}
\label{rho}
    \rho^{2} = \sum^{N}_{\alpha=1} |\psi_{\alpha}|^{2}
\end{equation}
        is nonvanishing.
        In this case it is natural to use
    $ \rho $
        as one of the independent variables and introduce new variables
\begin{equation*}
    \chi_{\alpha} = \psi_{\alpha} /\rho
\end{equation*}
        such that
\begin{equation}
\label{sph}
    \sum^{N}_{\alpha=1} |\chi_{\alpha}|^{2} = 1 .
\end{equation}

        In this way the compact target (I use the slang of the
        string theory)
    $ \SSS^{2N-1} $
        appears.

        When magnetic interaction is introduced the invariance
        with respect to the phase transformation
\begin{equation*}
    \psi_{\alpha}(x) \to e^{i\lambda(x)} \psi_{\alpha}(x)
\end{equation*}
        is invoked.
        This means that the target
    $ \SSS^{2N-1} $
    changes
\begin{equation*}
    \SSS^{2N-1} \to \SSS^{2N-1}/U(1) .
\end{equation*}
        In particular for
    $ N = 2 $
        we have
\begin{equation*}
    \SSS^{3}/U(1) \sim \SSS^{2}
\end{equation*}
        and the field
    $ \nn(x) $
        naturally appears.
        Quite satisfactorily the tensor
    $ H_{ik} $
        also emerges as a contribution to the magnetic field strength.

        Let us illustrate it in more detail.
        From the beginning we shall treat the stationary system, so
        no electric field will be used.

        The magnetic field is described in a usual way by means of the
        vector potential
    $ A_{k}(x) $
        and its interaction with
    $ \psi $-fields
    is introduced via covariant derivatives
\begin{equation*}
    \nabla_{k} \psi = \pr_{k} \psi + i A_{k} \psi .
\end{equation*}
        The energy density (of Landau-Ginsburg-Gross-Pitaevsky type) looks
        as follows
\begin{equation}
\label{LGGP}
    E = \sum^{2}_{\alpha=1} |\nabla_{k} \psi_{\alpha}|^{2}
        + \frac{1}{2} F^{2}_{ik} + V(|\psi_{\alpha}|) ,
\end{equation}
        where
\begin{equation*}
    F_{ik} = \pr_{i} A_{k} - \pr_{k} A_{i}
\end{equation*}
        is the field strength of the magnetic field.
        The energy is invariant with respect to the gauge transformations
\begin{equation*}
    \psi_{\alpha} \to e^{i\lambda} \psi_{\alpha} , \quad
        A_{k} \to A_{k} - \pr_{k} \lambda .
\end{equation*}
        We shall make the change of the field variables so that only
        gauge invariant ones will remain.
        For that observe that the first term in the RHS of
(\ref{LGGP})
    is a quadratic form in
    $ A $
\begin{equation*}
    \sum^{2}_{\alpha=1} |\nabla_{k} \psi_{\alpha}|^{2} =
    \sum^{2}_{\alpha=1} |\pr_{k} \psi_{\alpha}|^{2}
        + A_{k} J_{k} + \rho^{2} A^{2}_{k} ,
\end{equation*}
        where we use variable
    $ \rho $
        from
(\ref{rho})
        and introduce current
\begin{equation*}
    J_{k} = -i \sum^{2}_{\alpha=1}
        (\bar{\psi}_{\alpha} \pr_{k} \psi_{\alpha} -
            \pr_{k} \bar{\psi}_{\alpha} \psi_{\alpha}) .
\end{equation*}
        It is easy to check, that under the gauge transformations the
        current
    $ J_{k} $
        changes as follows
\begin{equation*}
    J_{k} \to J_{k} + 2 \rho^{2} \pr_{k} \lambda ,
\end{equation*}
        so that the sum
\begin{equation*}
    C_{k} = A_{k} + \frac{1}{2\rho^{2}} J_{k}
\end{equation*}
        is gauge invariant.
        We shall use this variable instead of
    $ A_{k} $.
        Another gauge invariant combination is given by the quadratic form
\begin{equation}
\label{himap}
    \nn = (\bar{\chi}_{1},\bar{\chi}_{1}) \, \vec{\tau}
        \begin{pmatrix}
            \chi_{1} \\ \chi_{2}
        \end{pmatrix} ,
\end{equation}
        where
    $ \vec{\tau} = (\tau_{1},\tau_{2},\tau_{3}) $
        are Pauli matrices
\begin{equation*}
    \tau_{1} = \begin{pmatrix}
            0 & 1 \\
            1 & 0
        \end{pmatrix} , \quad
    \tau_{2} = \begin{pmatrix}
            0 & -i \\
            i & 0
        \end{pmatrix} , \quad
    \tau_{3} = \begin{pmatrix}
            1 & 0 \\
            0 & -1
        \end{pmatrix} .
\end{equation*}
        Normalization
(\ref{sph})
        for
    $ \chi_{\alpha} $
        implies that
    $ \nn $
        is a real unit vector.
        In fact the map
\begin{equation*}
    (\chi_{1} , \chi_{2}) \to \nn
\end{equation*}
        defined in
(\ref{himap})
        is a standard Hopf map.
        Variable
    $ \nn $
        is manifestly gauge invariant and the set of variables
    $ (\rho, \nn, C_{k} ) $
        is our gauge invariant choice, substituting for the initial set
    $ (\psi_{\alpha}, A_{k} ) $.
        The energy density can be explicitly expressed via
    $ \rho, \nn $
        and
    $ C_{k} $
        as follows
\begin{equation*}
    E = (\pr_{k} \rho)^{2} + \rho^{2} ((\pr_{k} \nn)^{2} + C^{2}_{k}) +
        \frac{1}{2} (\pr_{k} C_{i} - \pr_{i} C_{k} + H_{ik})^{2} +
            v(\rho, n_{3}) .
\end{equation*}
        The most notable feature is the appearance of the tensor
    $ H_{ik} $,
        defined in
(\ref{HH}).
        The model, described in the main text, emerges if we put
    $ \rho = \text{const} $
        and
    $ C = 0 $.
        Hopefully nontrivial
    $ \rho $
        and
    $ C $,
        at least confined to some range, do not spoil the soliton picture.
        This problem is under discussion now, see
\cite{BFN},
\cite{PP}.

        Let us stress, that the use of two fields
    $ \psi_{\alpha} $,
    $ \alpha = 1,2 $
    is most essential in this
        example.
        If
    $ N = 1 $
        only variables
    $ \rho $
        and
    $ C $
        remain after the reduction, similar to just described.
        If
    $ N > 2$
        the
    $ CP(N-1) $
        field generalizing
    $ \nn $
    has no topological characteristics.

        Another application, considered recently
\cite{FNP},
        deals with the parameterization for the
    $ SU(2) $
        Yang-Mills field
    $ A^{a}_{\mu}(x) $,
    $ \mu = 0,1,2,3 $,
    $ a = 1,2,3 $.
        The Yang-Mills Lagrangian is invariant with respect to the
        nonabelian gauge transformations
\begin{equation*}
    \delta A^{a}_{\mu} =\pr_{\mu} \eps^{a} + f^{abc}  A^{b}_{\mu} \eps^{c}.
\end{equation*}
        However in some treatments one reduces this invariance
    by the partial gauge fixing to the abelian one
\begin{gather*}
    \delta B_{\mu} = i \eps B_{\mu} \, , \quad
    \delta A^{3}_{\mu} = \pr_{\mu} \eps ,
\end{gather*}
        where
    $ B_{\mu} = A^{1}_{\mu} + i A^{2}_{\mu} $
        is a complex vector field.
        I shall not discuss the reason for this reduction here
    and proceed assuming that it is done.
        Observe, that two vector fields
    $  A^{1}_{\mu}$, $ A^{2}_{\mu} $
        in generic situation define a plane in Minkowski space
        and introduce an orthonormalized basis in this plane
    $ \ee^{\alpha}_{\mu} $,
    $ \alpha = 1,2 $.
\begin{equation*}
    \ee^{\alpha}_{\mu} \ee^{\beta}_{\mu} = \delta_{\alpha \beta} .
\end{equation*}
        Let
    $ \ee_{\mu} = \ee^{1}_{\mu} + i \ee^{2}_{\mu} $.
        The basis is defined up to rotation
\begin{equation*}
    \ee_{\mu} \to e^{i\omega} \ee_{\mu}.
\end{equation*}
        The fields
    $ B_{\mu} $
        can be written in terms of this basis as
\begin{equation*}
    B_{\mu} = \psi_{1} \ee_{\mu} +  \psi_{2} \bar{\ee}_{\mu}
\end{equation*}
        and thus two complex valued fields
    $ \psi_{1} $ and
    $ \psi_{2} $
    appear.
        The situation becomes quite similar to the previous
        example and indeed in
\cite{FNP}
        the complete parameterization of the Yang-Mills variables is
        introduced with appearance of
    $ \nn $-field
    and corresponding
    $ H $-tensor.
        This is an indication that the Yang-Mills theory can have string-like
        excitations.
        However the situation is not that simple.
        The classical Yang-Mills theory is conformally invariant and
        has no dimensional parameters.
        Thus no hope for the localized regular classical solution exists.
        Nevertheless this complication could be lifted by quantum corrections.
        The famous ``dimensional transmutation'', which leads to the
    appearance of dimensional parameter in quantum effective action,
    could favor the nonvanishing value of the corresponding
    $ \rho $-variable.
        All these considerations at the moment are rather speculative and
        need much more work to become reasonable.
        Personally I am quite impressed by this possibility and continue
        to work on it.

\section{Conclusions}

\vskip-5mm \hspace{5mm}

        I think that the topic of my talk is quite instructive.
        It connects different domains in mathematics and mathematical
        physics: nonlinear PDE, elementary topology, quantum field
        theory, numerical methods.
        It illustrates the essential unity of mathematics, theoretical
        and applied.
        Finally it could lead to the realistic physical applications.
        For all these reasons I decided to present it to the ICM2002.

\label{lastpage}


\begin{thebibliography}{99}

\bibitem{KZ}
    M.~D.~Kruskal, N.~Zabusky,
    Interaction of ``Solitons'' in a Collisionless Plasma and
    the Recurrence of the Initial States.
    {\it Phys. Rev. Letters}, {\bf 15} (1965), 240--243.

\bibitem{ggkm}
    G.~S.~Gardner, J.~M.~Greene, M.~D.~Kruskal, R.~M.~Miura,
    Method for solving the Korteveg-de-Vries equation.
    {\it Phys. Rev. Lett.}, {\bf 19} (1967), 1095.

\bibitem{ZSh}
    V.~E.~Zakharov, A.~B.~Shabat, Exact theory of two-dimensional
    self-focusing and one-dimensional self-modulation of waves
    in nonlinear media, {\it Soviet Phys. JETP}, {\bf 34} (1972), 62--69.

\bibitem{FT}
    L.~D.~Faddeev, L.~A.~Takhtajan,
    {\it Hamiltonian Methods in the Theory of Solitons},
    Springer-Verlag Berlin Hiedelberg 1987.

\bibitem{LH}
    L.~D.~Faddeev,
    {\it How algebraic Bethe Ansatz works for integrable models},
    Proc. of Les Houches summer school, session LXIV, 149--220,
    NATO ASI, Elsevier 1998.

\bibitem{princeton}
    L.~D.~Faddeev,
    Quantization of Solitons.
    {\it Preprint IAS} print-75-QS70, 1975.

\bibitem{skyrme}
    T.~H.~R.~Skyrme,
    A Nonlinear Field Theory,
    {\it Proc. Roy. Soc. London}, {\bf A260} (1969), 127--138.

\bibitem{HP}
    G.~'t~Hooft,
    Magnetic Monopoles in Unified Gauge Theories,
    {\it Nucl.Phys.}, {\bf B79} (1974), 276--284. \\
    A.~M.~Polyakov,
    Particle Spectrum in the Quantum Field Theory.
    {\it Pisma Zh.Eksp.Teor.Fiz.}, {\bf 20} (1974), 430--433 (in Russian),
    {\it JETP Lett.} {\bf 20} (1974), 194--195.

\bibitem{FNN}
    L.~D.~Faddeev, A.~Niemi,
    Knots and Particles,
    {\it Nature}, {\bf 387} (1997), 58.

\bibitem{HS}
    J.~Hietarinta, P.~Salo,
    Faddeev-Hopf Knots: Dynamics of Linked Unknots.
    {Phys. Lett.}, {\bf B451} (1999), 60--67.

\bibitem{BS}
    R.~Battye, P.~M.~Sutcliffe,
    Knots as Stable Soliton Solutions in a Three-Dimensional Classical
    Field Theory.
    {\it Phys. Rev. Lett.}, {\bf 81} (1998), 4798--4801.

\bibitem{FNP}
    L.~D.~Faddeev, A.~Niemi,
    Aspects of Electric-Magnetic Duality in SU(2) Yang-Mills Theory.
    {\it Phys. Lett.}, {\bf B525} (2002), 195--200.

\bibitem{BFN}
    E.~Babaev, L.~D.~Faddeev, A.~Niemi,
    Hidden Symmetry and Duality in a Charged Two Condensate Bose System.
    {\it Phys. Rev.}, {\bf B65} (2002), 100512.

\bibitem{MM}
    M.~I.~Monastyrsky,
    {\it Topology of Gauge Fields and Condensed Matter},
    Plenum, New York, USA, 1993.

\bibitem{VK}
    A.~F.~Vakulenko, L.~V.~Kapitansky,
    Stability of Solitons in $ S^2 $ Nonlinear
    $ \sigma$-model,
    {\it Doklady of Soviet Acad. Sci.}, {\bf 246} (1979), 840 (in Russian),
    English translation in {\it Sov. Phys. Dokl.}, {\bf 24} (1979), 433.

\bibitem{RR}
    Yu.~P.~Rybakov,
    Structure of Minimizators of Energy in $ S^2 $ Nonlinear
    $ \sigma$-model.
    {\it Vestnik of Lumumba Univ. (PUDN), sec. ``Mathematika''} {\bf 2}
    (1995) 35--41.

\bibitem{HH}
    J.~Hietarinta,
    See page in http://users.utu.fi/hietarin/knots/index.html

\bibitem{PP}
    A.~P.~Protogenov,
    Charge Density Bounds in Superconducting States of Strongly
    Correlated Systems,
    {\it e-Print Archive:} cond-mat/0205133.

\end{thebibliography}
\end{document}